\documentclass[iop]{emulateapj}

\usepackage[T1]{fontenc}
\usepackage{amsmath}
\usepackage{natbib}
\shorttitle{Electron tail in afterglows}
\shortauthors{J{\'o}hannesson \& Bj{\"o}rnsson}

\begin{document}

\title{Low-energy electrons in GRB afterglow models}

\email{gudlaugu@hi.is, gulli@hi.is}

\author{Gu{\dh}laugur J{\'o}hannesson}
\affiliation{Science Institute, University of Iceland, Dunhagi 3, 107 Reykjav{\'i}k, Iceland}
\affiliation{Nordita
KTH Royal Institute of Technology and Stockholm University
Roslagstullsbacken 23, SE-106 91 Stockholm, Sweden}

\author{Gunnlaugur Bj{\"o}rnsson}
\affiliation{Science Institute, University of Iceland, Dunhagi 3, 107 Reykjav{\'i}k, Iceland}

\newcommand\nar{NewAR}

\begin{abstract}

   Observations of gamma-ray burst (GRB) afterglows have long provided the
   most detailed information about the origin of this spectacular phenomena.
   The model that is most commonly used to extract physical properties of the
   event from the observations is the relativistic fireball model, where
   ejected material moving at relativistic speeds creates a shock wave when it
   interacts with the surrounding medium. Electrons are accelerated in the
   shock wave, generating the observed synchrotron emission through
   interactions with the magnetic field in the downstream medium. It is usually assumed
   that the accelerated electrons follow a simple power-law distribution in
   energy between specific energy boundaries and that no electron exists outside
   these boundaries. This work explores the consequences of adding a
   low-energy power-law segment to the electron distribution whose energy
   contributes insignificantly to the total energy budget of the distribution. The low-energy
   electrons have a significant impact on the radio emission, providing
   synchrotron absorption and emission at these long
   wavelengths. Shorter wavelengths are affected through the normalization of the distribution. The
   new model is used to analyze the light curves of GRB 990510 and the
   resulting parameters compared to a model without the extra electrons. The
   quality of the fit and the best fit parameters are significantly affected by the additional model
   component. The new component is in one case found to strongly affect
   the X-ray light curves showing how changes to the model at radio
   frequencies can affect light curves at other frequencies through changes
   in best fit model parameters.

\end{abstract}

\keywords{gamma-ray burst: general --- methods: data analysis --- radiation
mechanisms: non-thermal}

\section{Introduction} \label{sec:intro}

Gamma-ray bursts (GRBs) are the most powerful explosions in the Universe and can
therefore be observed to very high redshift. They have
been hypothesised to be tracers of star formation and thus be a 
probe of the star formation history of the early universe
\citep{CharyEtAl:2016}.  Using the bursts
as effective tools in cosmological studies requires a solid understanding of
the physics that drives the explosions and the observable consequences of the
GRB-events \citep{WangEtAl:2015}. Our understanding of GRBs comes mostly through
observations of their afterglow emission at wavelengths ranging from radio to
X-rays \citep[e.g.][]{Piran:2004,GehrelsEtAl:2009}. These observations are
best interpreted with
a model where the emission arises from shocks in a relativistically
expanding jet, internal shocks for the prompt high-energy GRB emission and
external shocks for the afterglow.  Electrons are accelerated to high energies
in these shocks giving rise to synchrotron emission as they interact with the
magnetic field in the downstream medium.

The theory of particle acceleration in relativistic shocks is far from
complete but it is generally acknowledged that a population of high-energy
particles form whose distribution can be approximated as a power-law
in momentum \citep{PelletierEtAl:2017}. These
high energy particles contain most of the energy of the distribution and are
responsible for the non-thermal emission arising in the model. There
are, however, a considerable number of lower energy electrons that can
contribute to the long-wavelength emission and increase absorption.
\citet{ResslerLaskar:2017} studied the effect of adding a thermal population
of electrons to the power-law distribution and find that it can provide
significant effects even in the optical frequency range depending on the
chosen parameters. Their analysis showed calculated light curves for several
models but they did not compare it to any observations to test the validity
of the model.

This work also focuses on the effects of low-energy electrons by extending the
electron distribution to low energies using a power-law segment. 
A power-law distribution is chosen over a thermal distribution for simplicity to 
demonstrate the effect on parameter determination from fitting afterglow observations.
A thermal distribution requires at least two parameters and special care to get a 
continuous distribution while the extra power-law segment requires only a single 
parameter and the distribution is automatically continuous.
The
extension is added to the GRB afterglow model of \citet{JohannessonEtAl:2006}
that has been used to analyze several GRB afterglows
\citep{deUgartePostigoEtAl:2005,deUgartePostigoEtAl:2007,ResmiEtAl:2012,Sanchez-RamirezEtAl:2017}.
To test the effect of the additional electrons the model 
is used to analyse the afterglow observations of GRB
990510 that has been well studied
\citep[e.g.][]{PanaitescuKumar:2001,PanaitescuKumar:2002,JohannessonEtAl:2006}.  
The results show a statistical preference for models with additional low-energy
electrons indicating the need for further exploration of the electron
distribution in GRB afterglow models.
The paper is organized as follows: in
section~\ref{sec:model} the model and the new extension is described and
its effects explored, in
section~\ref{sec:analysis} the results of the analysis of GRB~990510 are presented and the
paper concludes with discussion in section~\ref{sec:conclusions}.

\section{Model} 
\label{sec:model}

The model in \citet{JohannessonEtAl:2006} assumes the afterglow emission
arises from a relativistic shock wave traveling through the central engine's
surrounding medium. The shock wave is formed as a relativistic slab of matter
with energy $E_0 = \Gamma_0 M_0 c^2$ is released into a cone having a half opening
angle $\theta_0$. The shock
is assumed to accelerate electrons to relativistic speeds and a strong
magnetic field is generated within the downstream medium resulting in synchrotron emission. The
dynamics of the system is determined from energy and momentum
conservation, assuming the downstream medium is a thin uniform shell that expands
sideways at the local speed of sound and sweeps up everything in its way
\citep{Rhoads:1999}. 
The density of the surrounding medium can have an arbitrary radial dependence
but in this work only two forms are considered, a constant density
medium, $\rho(r) = m_p n_0$, and a wind-like medium, $\rho(r) = C_w A_* r^{-2}$.
Here $r$ is the distance from the central engine, $m_p$ is the mass of the
proton, $n_0$ is the number density of the external medium, $A_*$ is a normalization
parameter, and $C_w = 5.015\cdot10^{11}\,{\rm g\; cm^{-1}}$ for a typical
Wolf-Rayet star \citep{DaiLu:1998, ChevalierLi:1999}.

The electron energy distribution in this model is based on the one described
in \citet{PanaitescuKumar:2001}, but extended to low energies with an
additional power-law section.  The electron distribution is
\begin{equation}
   \frac{d n}{d\gamma} = n_{e,0}
   \begin{cases}
      \left( \frac{\gamma}{\gamma_1} \right)^{-p_l}, & \gamma_0 < \gamma < \gamma_1, \\
      \left( \frac{\gamma}{\gamma_1} \right)^{-p_c}, & \gamma_1 < \gamma <
      \gamma_2, \\
      \left( \frac{\gamma_2}{\gamma_1} \right)^{-p_c}\left(
      \frac{\gamma}{\gamma_2} \right)^{-p-1}, & \gamma_2 < \gamma < \gamma_M ,
   \end{cases}
   \label{eq:elDistribution}
\end{equation}
where $\gamma_1 = \min\left\{ \gamma_i, \gamma_c \right\}$, $\gamma_2 =
\max\left\{ \gamma_i, \gamma_c \right\}$,  $p_c = p$ in
the slow cooling phase where
$\gamma_i < \gamma_c$, and $p_c = 2$ in the fast cooling phase with $\gamma_c <
\gamma_i$.  The lower limit of the distribution is fixed at $\gamma_0 = 2$ because 
our formalism for the emitting radiation is only valid for high-energy electrons.  
Here, $\gamma_i$ is the injection break Lorentz
factor which is defined assuming the electrons at the injection break contain
a fraction $\epsilon_i$ of the total kinetic energy of the downstream medium,
\begin{equation}
   \gamma_i = \epsilon_i \frac{m_p}{m_e}(\Gamma - 1) + 1,
   \label{eg:gammaInjection}
\end{equation}
where $m_e$ is the mass of the
electron.  The cooling Lorentz factor is,
\begin{equation}
   \gamma_c = \frac{4 \pi m_e c}{\sigma_T B'^2 t' },
   \label{eq:coolingBreak}
\end{equation}
which is found by equating the synchrotron energy loss at time $t'$ with the
energy of the electrons \citep{Kardashev:1962}.  Primed quantities are evaluated
in the co-moving rest-frame of the shock wave.  Here $\sigma_T$ is the
Thompsons' cross section and $B' =
\sqrt{32\pi \epsilon_B \Gamma\left( \Gamma -1 \right) \rho c^2}$ is the
magnetic field strength.

The energy of the distribution should be dominated by electrons with Lorentz
factors above or around $\gamma_1$ so $p_l < 2$. For large {\em negative} values of $p_l$, 
the distribution behaves effectively as
a distribution without the low energy extension.
The normalization factor $n_{e,0}$ is determined from particle conservation.
The maximum Lorentz factor $\gamma_M$ is 
determined such that the acceleration timescale does not exceed the radiative
loss timescale \citep{DaiLu:1998} and the total energy of the electron distribution does not
exceed a fraction $\epsilon_e < 1$ of the kinetic energy of the downstream medium.  Depending on
the exact values of the parameters, the latter condition can result in a
sharp break in the emitted spectrum above the synchrotron frequency associated
with $\gamma_M$.  This break can even extend down to 
the optical range or lower at late times for certain parameter values.

The synchrotron radiation is calculated using the standard assumption that the pitch angle between the
electrons and the magnetic field is isotropic. The radiation is calculated
numerically in the model by integrating the synchrotron power per electron over the electron distribution.  Using standard
assumptions \citep[e.g.][]{SariEtAl:1998} one can easily derive an approximate
power-law behavior for the resulting co-moving frame radiation power
\begin{equation}
   P'_\nu(\nu') \propto
   \begin{cases}
      \nu'^{1/3}, & {\rm if} \quad \nu' < \nu'_l, \\
      \nu'^{-(p_l - 1)/2}, & {\rm if} \quad \nu'_l \le \nu' < \nu'_1, \\
      \nu'^{-(p_c - 1)/2}, & {\rm if} \quad \nu'_1 \le \nu' < \nu'_2, \\
      \nu'^{-p/2}, & {\rm if} \quad \nu'_2 \le \nu' < \nu'_M. \\
   \end{cases}
   \label{eq:powerLawApprox}
\end{equation}
Here $\nu'_k$ is the synchrotron frequency corresponding to the
Lorentz factor $\gamma_k$. It is assumed that $p_l > 1/3$. 
The additional low-energy electrons therefore
contribute to the emitted spectrum for $\nu' < \nu'_1$. The contribution
to the synchrotron absorption coefficient can also be evaluated and if $p_l >
-2/3$ then
\begin{equation}
   \alpha'_\nu(\nu') \propto
   \begin{cases}
      \nu'^{-5/3}, & {\rm if} \quad \nu' < \nu'_l, \\
      \nu'^{-(p_l + 4)/2}, & {\rm if} \quad \nu'_l \le \nu' < \nu'_1, \\
      \nu'^{-(p_c + 4)/2}, & {\rm if} \quad \nu'_1 \le \nu' < \nu'_2, \\
      \nu'^{-(p+5)/2}, & {\rm if} \quad \nu'_2 \le \nu' < \nu'_M. \\
   \end{cases}
   \label{eq:tauPowerLawApprox}
\end{equation}
The low energy electrons contribute to the absorption of the
synchrotron spectrum even in the case their contribution to the emission is
insignificant.  In addition to the change in spectrum, the low-energy
electrons affect the estimation of the normalization constant $n_{e,0}$ and
therefore the normalization of the emission at all frequencies.  Numerical calculations
show that $p_l \lesssim -10$ is required before the effects of the extra component can be completely
neglected although the effect is small up
to $p_l \lesssim -2$. 

\begin{figure}[tb]
   \centering
   \includegraphics[width=0.48\textwidth]{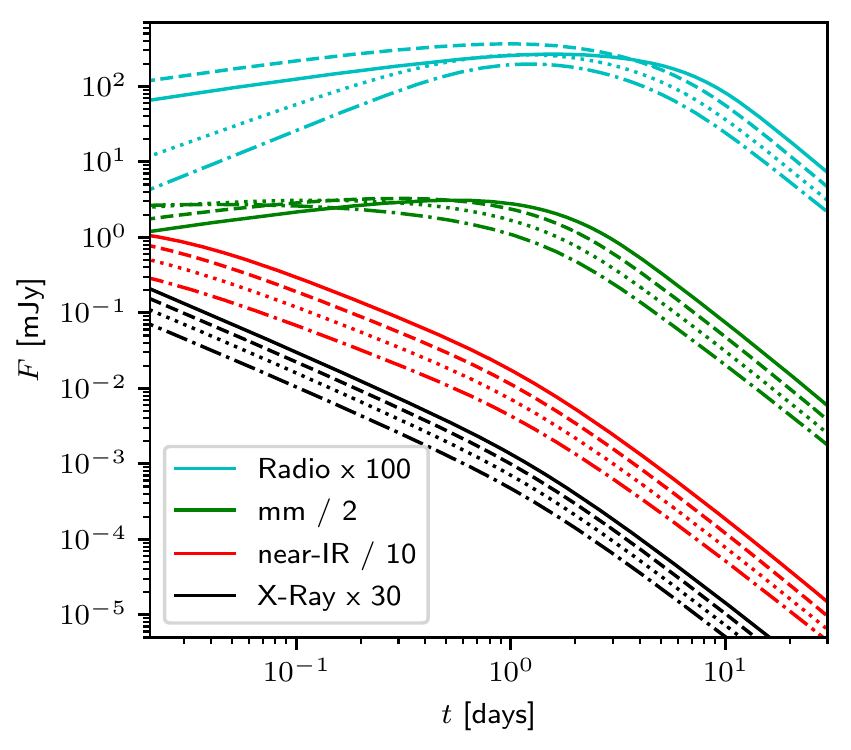}
   \caption{Resulting light curves from example calculations were the value of
   $p_l$ is varied. Four frequency bands are shown: radio (5~GHz, cyan), mm
 (100~GHz, green), near-IR (K-band, red), and X-rays (3~keV, black). 
 Four values
 of $p_l$ are used: -10 (solid curve), -0.66 (dashed curve), 0.33 (dotted
 curve), and 0.75 (dash-dotted curve).}
   \label{fig:exampleCurves}
\end{figure}

The effect of the low-energy extension is illustrated in
Figure~\ref{fig:exampleCurves} that shows results of model calculations for
several values of $p_l$. The other parameters are fixed at typical values for
GRB afterglows: $E_0=10^{51}$~erg, $\Gamma_0=1000$, $\theta_0 = 3^\circ$, $n_0
= 1$~cm$^{-3}$, $p = 2.2$, $\epsilon_e=0.3$, $\epsilon_i=0.01$, and
$\epsilon_B=0.001$. As expected from the analytical
approximations in Eq.~(\ref{eq:powerLawApprox}) the effect is largest for the
radio and mm light curves whose frequencies are below $\nu_i$ before 10
and 1 days, respectively.  The radio light curve is particularly affected
because of increased synchrotron self-absorption for $p_l > 0.33$ that shows
up in the fast rise of the early radio light curve.  The effect on the near-IR
to X-rays is in this case only through the normalization of the light curve.
The increased number of electrons at low energies results in lower emission,
even though the total energy of the electron distribution is dominated by
electrons around $\gamma_i$.

\section{Application to GRB 990510} 
\label{sec:analysis}

The afterglow of GRB 990510 has been referred to as the canonical afterglow
light curve due to its smooth decline that can be well modeled with
synchrotron emission from a collimated shock wave \citep{StanekEtAl:1999,
   HarrisonEtAl:1999, KuulkersEtAl:2000, HollandEtAl:2000, PanaitescuKumar:2001,
JohannessonEtAl:2006}. Its afterglow is well sampled with data at many
wavelengths from radio to X-rays. The data used in this analysis is 
from \citet{HarrisonEtAl:1999} (optical and radio), \citet{StanekEtAl:1999}
(optical), and \citet{KuulkersEtAl:2000} (X-rays).  The optical data is
corrected for Milky-Way dust extinction of E(B-V)=0.2 \citep{SchlegelEtAl:1998}.

The Bayesian method is used to test the effect of the additional low-energy
electrons by comparing the Bayesian factor of models with and without the
low-energy component.  
Models with both a constant density medium (CM, CMo) and a
wind-like medium (WM, WMo) are used.  The 'o' in the
model names stands for without the low-energy electrons.
The Bayesian evidence is calculated using MultiNest,
which also provides posterior distributions for the free parameters of the
model \citep{FerozHobson:2008,FerozEtAl:2009,FerozEtAl:2013}.  The likelihood
is calculated assuming the afterglow flux data are sampled from a log-normal
distribution which is equivalent to the apparent magnitude being distributed
normally.  The prior distributions of the parameters are mostly
non-informative uniform or log-uniform distributions bounded only by physical
constraints of the model.  $\theta_0$ is bound from above to be no larger than
$90^\circ$, $\epsilon_e$ and $\epsilon_B$ are constrained to be less than
0.5, and $\epsilon_i$ is then constrained to be less than $\epsilon_e$.  The
initial energy release, $E_0$, is constrained to be less than
$10^{52}$~erg which is about 10\% of the energy expected to be released in the
gravitational collapse of a massive star.  The value of $p_l$ is also
constrained to be less than 1 so both the energy and number of electrons in
the distribution peaks at around $\gamma_i$.  Other boundaries are set such
that they do not affect the results.

\begin{figure}[tb]
   \centering
   \includegraphics[width=0.48\textwidth]{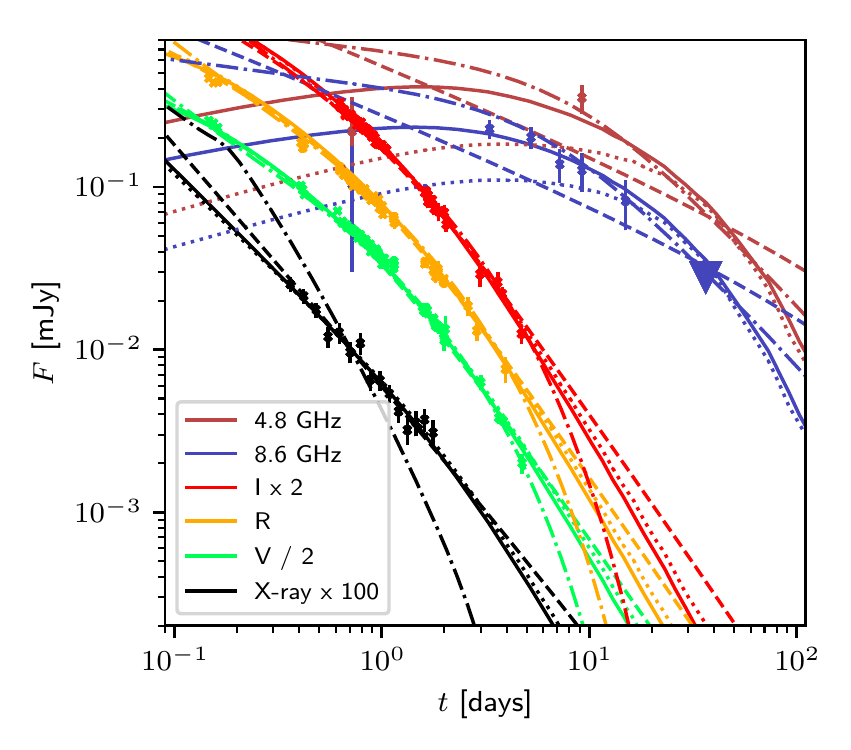}
   \caption{Afterglow light curves of GRB 990510. Data are shown with error
   bars overlaid with best fit models as curves. CM is
solid, CMo is dotted, WM is dashed, and WMo is dash-dotted.  Different
colors represent different wavebands as illustrated in the legend.}
   \label{fig:990510lcs}
\end{figure}

\begin{figure*}[tb]
   \centering
   \includegraphics[width=\textwidth]{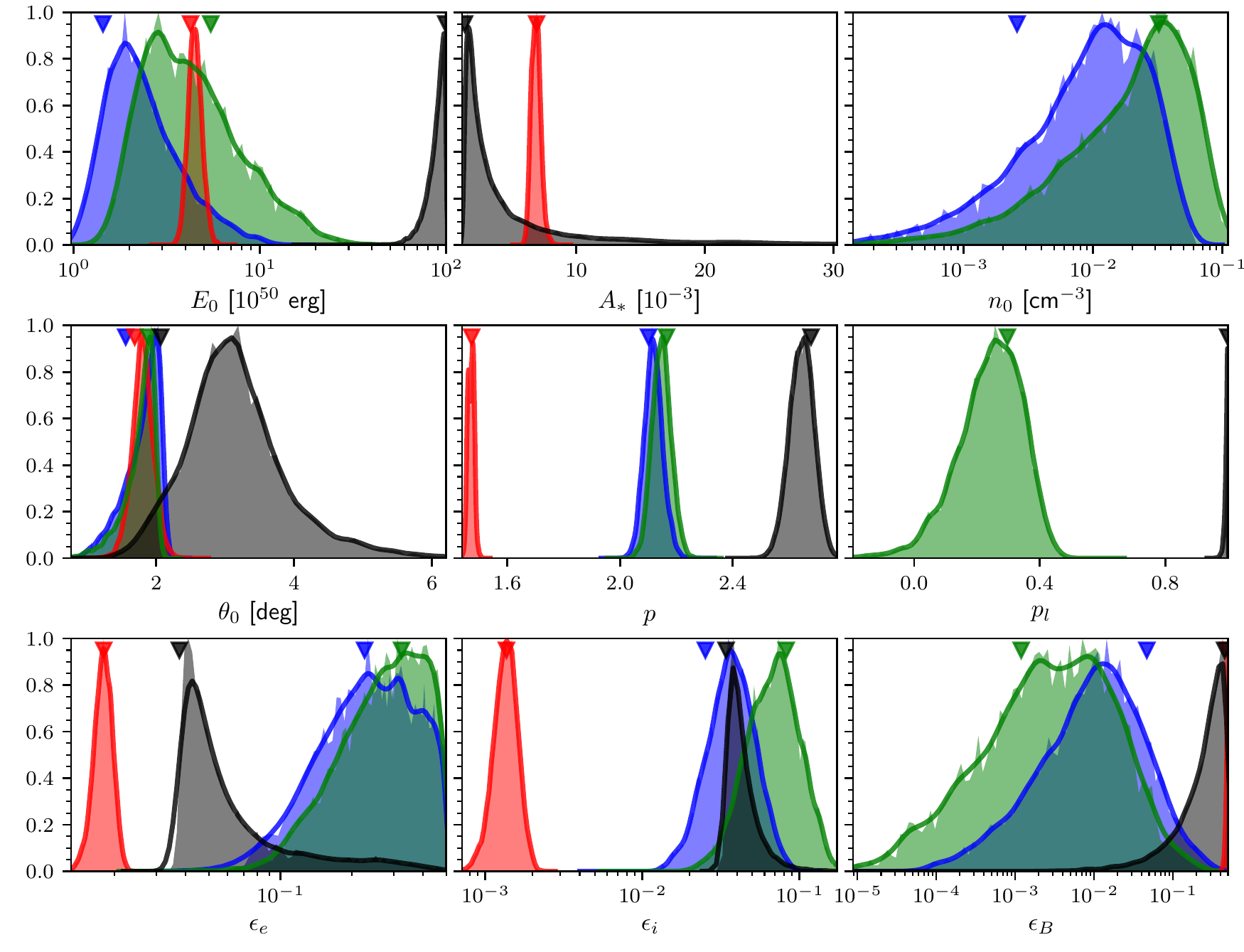}
   \caption{Marginalized posterior distributions for the parameters of the
      afterglow tuned to the data of 990510 shown in
      Figure~\ref{fig:990510lcs}.  The distributions are normalized to 1 at
      the peak. Shown are both the actual sampled
   distributions as half-transparent colors and the Gaussian kernel density
estimate as smooth curves. CM is green, CMo is blue, WM is black, and WMo is
red. The best fit values used
for the model curves in Figure~\ref{fig:990510lcs} are shown as symbols at the
top of the panels.  Note that the best fit values are not always near the peak of the distribution.}
   \label{fig:990510post}
\end{figure*}

\begin{deluxetable*}{lcccc}
   \tablecaption{\label{tab:990510posterior} Resulting posterior mean values with the
 associated 68\% confidence intervals for the model parameters. Number in
 parentheses is the value associated with the maximum likelihood used to
 create the model curves in Figure~\ref{fig:990510lcs}.}
   \tablecolumns{5}
   \tablehead{\colhead{Parameter} & \colhead{CMo} & \colhead{WMo} & \colhead{CM} & \colhead{WM} }
   \startdata
   Bayesian evidence & $-317.3 \pm 0.4$  & $-454.81 \pm 0.03$  & $-280.01 \pm 0.03$  & $-265.6 \pm 0.2$ \\
   Minimum $\chi^2$ & $549.8$  & $814.7$  & $470.1$  & $444.1$ \\
   \hline
   $E_0$ [$10^{50}$ erg] & $2.65^{+1.5}_{-0.66}$  ($1.44$) & $4.54^{+0.41}_{-0.35}$  ($4.25$) & $5.3^{+4.1}_{-1.6}$  ($5.5$) & $89.5^{+6.0}_{-11}$  ($99.6$)\\
   $A_*$ [$10^{-3}$] & \nodata & $6.87^{+0.36}_{-0.36}$  ($6.91$) & \nodata & $2.83^{+1.7}_{-0.50}$  ($1.31$)\\
   $n_0$ [cm$^{-3}$] & $0.0134^{+0.015}_{-0.0075}$  ($0.0026$) & \nodata & $0.028^{+0.029}_{-0.018}$  ($0.032$) & \nodata\\
   $\theta_0$ [deg] & $1.79^{+0.17}_{-0.34}$  ($1.56$) & $1.80^{+0.15}_{-0.15}$  ($1.69$) & $1.75^{+0.13}_{-0.28}$  ($1.89$) & $3.15^{+0.70}_{-0.60}$  ($2.07$)\\
   $p$ & $2.119^{+0.032}_{-0.032}$  ($2.101$) & $1.471^{+0.011}_{-0.012}$  ($1.474$) & $2.148^{+0.034}_{-0.034}$  ($2.165$) & $2.645^{+0.042}_{-0.044}$  ($2.678$)\\
   $p_l$ & \nodata & \nodata & $0.241^{+0.091}_{-0.11}$  ($0.297$) & $0.9943^{+0.0030}_{-0.0062}$  ($0.9989$)\\
   $\epsilon_e$ & $0.30^{+0.13}_{-0.12}$  ($0.23$) & $0.0181^{+0.0016}_{-0.0016}$  ($0.0181$) & $0.29^{+0.13}_{-0.11}$  ($0.32$) & $0.0757^{+0.050}_{-0.0094}$  ($0.0376$ )\\
   $\epsilon_i$ & $0.039^{+0.015}_{-0.011}$  ($0.025$ ) & $0.00137^{+0.00024}_{-0.00023}$  ($0.00137$) & $0.073^{+0.031}_{-0.024}$  ($0.082$) & $0.0478^{+0.015}_{-0.0058}$  ($0.0342$)\\
   $\epsilon_B$ & $0.0232^{+0.031}_{-0.0086}$  ($0.0467$) & $0.4818^{+0.0096}_{-0.021}$  ($0.4971$) & $0.0103^{+0.015}_{-0.0026}$  ($0.0012$) & $0.28^{+0.14}_{-0.17}$  ($0.45$)\\
   \enddata
\end{deluxetable*}

The best fit model is the WM model where the logarithm of the Bayesian evidence is $\log(Z) =
-266$. This is considerably higher than the CM model which has $\log(Z) =
-280$ giving a value of 10 for the log of the Bayes factor indicating strong
evidence.  The WM model
gives a better fit to the R-band data while the radio data is better fit with the
CM model.  Other bands are similar for the models and the WM model thus
provides a larger Bayesian evidence because the number of points in the R
band is much larger than that in radio.
The WMo and CMo models result in significantly worse Bayesian evidence than
the corresponding WM and CM models.  The Bayesian factor between the CM and
CMo models is 37 and for WM and WMo it is 189, providing very strong
evidence for the addition of the low energy electrons in this analysis.

The best fit model curves are shown overlaid on the data in
Figure~\ref{fig:990510lcs} for all models.  Even though the WM model is
statistically better, the CM model looks better because it follows the trend
of the radio points and the difference in the R-band is barely visible.  It is
also clear that the WMo model provides a poor
fit to the X-ray data.  The low-energy electrons thus significantly affect the
quality of the fit at X-ray wavelengths through changes in the model
parameters even though they do not contribute to
the emission at those wavelengths.  It also demonstrates how important multi-wavelength data is
for model selection.

The 1D marginal posterior distributions
for the model parameters are shown in Figure~\ref{fig:990510post} for
the models considered in the analysis.  
Table~\ref{tab:990510posterior} shows the parameters posterior mean values, 
the 68\% confidence regions, and the maximum likelihood values.
The WM model clearly stands out and 
the posterior distributions are often cut off
abruptly by the prior range.  The WM model requires large values
for the initial energy release, $E_0$, which is only constrained by the
selection of prior.  Increasing the size of the
prior results in the best fit model having even higher values of $E_0$ which
are beyond reasonable estimates of the available energy from the central
engine.  The posterior distribution for $p_l$ is also at the boundary set by
the prior. Because of these extreme parameter values and the fact that the CM
model better reproduces the radio data the CM model is considered a better
model of this event even though the WM model is statistically favored.

There is considerable difference
between the posterior distributions of the models with and without the
low-energy electrons, in particular for the WM and WMo models.  The
posterior
distributions for the WM and CM models are broader
compared to the WMo and CMo counterparts and the means of the distributions
are also shifted.
The effect differs somewhat between the CM and WM models in detail, but the shift is
in all cases in the same direction except for the external density.  There is
a decrease in the posterior mean for $A_*$ in the WM model compared to the
WMo model while $n_0$ is increased in
the CM model compared to the CMo model.  The 2D marginal posterior distributions (not shown) indicate
that $p_l$ is in both models well constrained and there is very little
correlation between $p_l$ and the other model parameters.

\section{Discussion and Summary} 
\label{sec:conclusions}

The best fit model parameters presented here are in reasonable agreement
with previous analysis.  \citet{PanaitescuKumar:2001} used a CMo model very
similar to the one used in this analysis.  Their best
fit values are all within the 99\% confidence intervals of our posterior
distributions apart from the value of $\theta_0$.  Their value for $\theta_0$
is $2.7^\circ$ which is considerably off the posterior distribution determined
here. A follow up study was performed in \citet{PanaitescuKumar:2002} where
the parameters changed significantly.  In particular their value of $\theta_0$
and $n_0$ are larger and both outside the posterior distributions presented in
Figure~\ref{fig:990510post}.  Their value of $p$ is also considerably smaller,
but with large uncertainties and agrees with the posterior at the
$2\sigma$~level. Like the present analysis, \citet{PanaitescuKumar:2002} find
that a constant density external medium better fits the data. This afterglow
data was also analysed by \citet{JohannessonEtAl:2006} using an older version
of the code used here and a different fitting technique. As
expected, the parameter estimates all fall within the 68\% confidence
intervals of the current posterior distributions.

The results of the analysis of GRB 990510 and the statistical preference for
the additional electron component lend support to the need for more detailed
treatment of the electron distribution in GRB afterglow modeling.  This can in
particular affect the determination of the energetics of the outflow and the
density structure of the external medium.  The
power-law segment added in this work is just a simple modification of the
electron energy distribution to explore
its effects and further work is needed to get a more accurate physical picture.
One such method is to simultaneously solve for the dynamics of the afterglow
and the distribution of electrons. This was done in the work of
\citep{GengEtAl:2018} but their approach is limited to electron cooling and lacks
the thermalization effect of the electrons that may be important for the
lowest energy electrons.  They also excluded several important effects, such
as the EATS. Clearly, there is room for considerable improvements in this
area of GRB afterglow modeling.




\end{document}